\documentclass[aps,showpacs,floatfix,amsmath,amssymb]{revtex4}
\usepackage{epsfig} % Include figure files two-column,

\begin{document}

\title{Eta-nucleon coupling constant in QCD with SU(3) symmetry breaking}
\author{Janardan P. Singh}
\affiliation{Physics Department, Faculty of Science, The M. S.
University of Baroda, Vadodara-390002, India}
\author{Frank X. Lee and Lai Wang}
\affiliation{Physics Department,
The George Washington University,  Washington,  DC 20052,  USA}

\begin{abstract}

We study the $\eta$NN coupling constant using the method of QCD sum rules
starting from the vacuum-to-eta correlation function of the
interpolating fields of two nucleons. The matrix element of this
correlation has been taken with respect to nucleon spinors to
avoid unwanted pole contribution. The SU(3)-flavor symmetry breaking effects have
been accounted for via the $\eta$-mass, s-quark mass and eta decay
constant to leading order. Out of the four sum rules obtained by
taking the ratios of the two sum rules in conjunction with the two
sum rules in nucleon mass, three are found to give mutually
consistent results. We find the SU(3) breaking effects significant, 
as large as 50\% of the SU(3) symmetric part.

\end{abstract}

 \vspace{1cm}  % Quarks, gluons, and QCD in nuclear reactions
%\keywords{QCD sum rule method, coupling constant, SU(3) flavor symmetry breaking}

\maketitle

%%%%%-------------------------------------------------------------------
\section{Introduction}
\label{intro}
The knowledge of eta-nucleon coupling constant, $g_{\eta N N}$,
has implications in both hadronic physics as well as nuclear
physics. It is required for study of $\eta$ production off a
nucleon target, and also for analysis of NN scattering data. In
general, it has been used for construction of realistic nuclear
potentials~\cite{Machleidt,Ceci,Stoks}. In particular,
$\eta$-exchange along with pion exchange between nucleons can give
rise to isospin violation in nuclear potential~\cite{Halperin};
this has been used for the resolution of the Nolen-Schiffer
anomaly~\cite{Nolen}. $g_{\eta N N}$ will also be useful for
understanding the strange content of the nucleon~\cite{Ellis} via
the Sullivan process~\cite{Sullivan}. It may provide important
constraints on parton distribution amplitudes (DAs) of $\eta$. For
mesons $\pi^0$, $\eta$ and $\eta'$, the strong interaction induces
transitions between quarks of different flavors ($u\bar u$,
$d\bar d$, $s\bar s$) or gluons (gg, ...)~\cite{Feldmann}. The
mixing phenomenon is strongly connected with the U(1)$_A$ anomaly of
QCD. Hence, it is expected that a reliable determination of
$g_{\eta N N}$ would shed considerable light on the U(1)$_A$
dynamics of QCD~\cite{Shore}. The measurement of $g_{\eta N N}$ is
a formidable task since the production of $\eta$ mesons from
single nucleons is dominated by the resonance $S_{11}$(1535)
irrespective of the probe~\cite{Hanhart}. Therefore, a reliable
theoretical estimate of $g_{\eta N N}$ is desirable. Among the
various methods used for calculating hadronic parameters, QCD sum
rules are especially useful. This method has already been used for
calculating several meson-baryon
couplings~\cite{Reinders85,Reinders84,Birse,Doi00,Kim00,Kim99,Kondo02,Kondo03,Doi04,Aliev}
and, in particular, $g_{\eta N N}$ in SU(3)-flavor symmetry limit
~\cite{Kim00,Doi00}. In this work we wish to apply this method to
calculate $g_{\eta N N}$ with leading SU(3)-flavor violating effect taken
into account.

QCD sum rules for calculation of meson-baryon coupling constant
was first used by Reinders \emph{et al}.~\cite{Reinders85,Reinders84} 
who first considered three-point correlation 
function for $g_{\pi NN}$. On finding of the inconsistency with
the Goldberger-Treiman (GT) relation they studied two-point
correlation function of interpolating fields of nucleons with the
pion in the initial state. In the former approach one had to
assume an extrapolation from large space-like momentum to zero
momentum for the meson whereas no such assumption was needed in
the latter. It was pointed out by Birse and Krippa~\cite{Birse}
that in the soft pion limit this latter sum rule can be obtained
from nucleon mass sum rule by chiral rotation, and hence it does
not constitute an independent sum rule from that for the nucleon
mass. Therefore, one must take finite meson momentum to arrive at
an independent sum rule.

The two-point correlation function with an initial meson state
gives rise to several independent sum rules which have been
studied in detail~\cite{Kim00,Doi00,Kim99,Kondo02}. It has been
observed~\cite{Kim00,Doi00} that the sum rule with the tensor
structure gives the most reliable result. An improvement of this
approach has been proposed by Kondo and Morimatsu~\cite{Kondo03}
wherein one takes matrix element of the correlation function with
respect to nucleon spinors. The sum rule obtained in this way is
free from ambiguity arising from the choice of the effective
interaction Lagrangian and the effect of nonzero mass of the meson
can also be taken into account.

The evaluation of two-point correlation function of nucleons
between vacuum and a one meson state requires knowledge of matrix
elements of nonlocal quark and gluon operators between vacuum and
one meson state. Such matrix elements have been extensively
studied in literature mainly for isospin-non-singlet members of
the octet pseudo-scalar family~\cite{Belyaev,Ball99,Ball}. There have
also been derivations, using various approaches, of
Gell-Mann-Okubo type of relations involving two-parton and
three-parton light-cone distribution functions of pions, kaons and
eta~\cite{Chen06,Kim08}. These relations can be used to get matrix
elements of nonlocal quark and gluon operators between vacuum and
one eta-meson state to leading order in SU(3)-breaking
corrections. It may be pointed out that constants appearing in
DAs, such as meson decay constants and others to be specified
below, enter both in the effective Lagrangian relevant for
low-energy physics and the light-cone wave functions utilized in
high energy reactions.

In Sec. II, we derive the sum rule giving necessary details for
the projected correlation function. In Sec. III, we analyze the
results numerically and discuss them. Finally, in Sec. IV we
summarize our work and give conclusion.

\section{FORMALISM AND CONSTRUCTION OF SUM RULE}
\label{sumrule}

We consider the correlator of the standard nucleon currents
between vacuum and one $\eta$-state:

\begin{equation}
\Pi(q,p)=i\int d^4 x e^{iqx} <0|T\{J_N(x),\bar J_N (0)\}|\eta(p)>,
\end{equation}
where  $J_N$ is the standard proton current
\begin{equation}
  J_N =\epsilon^{abc}[u^{aT}C\gamma_\mu u^b]\gamma_5 \gamma^\mu d^c,
\end{equation}
where a, b, c are color indices. The most general form of
$\Pi(q,p)$ is~\cite{Kondo03}
\begin{equation}
\Pi(q,p)=i\gamma_5 \hat p \Pi^{AV}+ i\gamma_5 \Pi^{PS}+ \gamma_5
\sigma^{\mu\nu} q_\mu p_\nu \Pi^T+ i\gamma_5 \hat q \tilde
\Pi^{AV}.
\end{equation}

The  $\eta$NN  coupling constant $g_{\eta N N}$ is defined through
the coefficient of the pole as~\cite{Kim99}
\begin{eqnarray}
&&\bar u (qr) \;(\hat q -m)\; \Pi(q,p) \;(\hat q -\hat p -m)
u(ks)|_{q^2=m^2,\; (q-p)^2=m^2}= i\lambda^2 g_{\eta N N} \bar
u(qr)\gamma_5 u(ks),
\end{eqnarray}
where $k=q-p$ and $u(qr)$ is a Dirac spinor with momentum $q$ and
spin $r$ and is normalized as
\begin{equation}
\bar u(qr)u(qr) = 2m.
\end{equation}
Following~\cite{Kondo03}, we define the projected correlation function
\begin{equation}
\Pi_+ (q,p) =\bar u(q)\gamma_0 \Pi(q,p) \gamma_0 u(q-p). \label{PI}
\end{equation}
$\Pi_+$ can be regarded as a function of $q_0$ in the reference
frame in which $\vec q=0$. We split the projected correlation
function into even and odd parts as
\begin{subequations}
\begin{equation}
\Pi_+^E (q_0^2) = \frac{1}{2} [\Pi(q_0)+\Pi(-q_0) ],\label{PE}
\end{equation}
\begin{equation}
\Pi_+^O (q_0^2) = \frac{1}{2q_0} [\Pi (q_0)-\Pi(-q_0) ].\label{PO}
\end{equation}
\end{subequations}
The dispersion relation satisfied by these functions are
\begin{subequations}
\begin{equation}
\Pi_+^E (q_0^2) = -\frac{1}{\pi} \int d {q_0}'\frac{{q_0}'}
{q_0^2-{q_0}^{'2}} Im\Pi_+({q_0}'),
\end{equation}
\begin{equation}
 \Pi_+^O (q_0^2) = -\frac{1}{\pi} \int d{q_0}'
\frac{1}{q_0^2-{q_0}^{'2}} Im\Pi_+({q_0}').
\end{equation}
\end{subequations}
We define Borel transform by $\hat B$
\begin{equation}
\hat B f(q_0^2) =
\lim_{n\to\infty,-q_0^2\to\infty,-\frac{q_0^2}{n}=M^2}
\frac{{(-q_0^2)}^{n+1}}{n!} {(\frac{d}{dq_0^2})}^n f(q_0^2),
\end{equation}
so that the dispersion relations become
\begin{subequations}
\begin{equation}
\hat B [\Pi_+^E (q_0^2)] = \frac{1}{\pi} \int d{q_0}' {q_0}'
exp(-\frac{{q_0}'2}{M^2}) Im\Pi_+ (q_0^{'}), \label{BE}
\end{equation}
\begin{equation}
\hat B [\Pi_+^O (q_0^2)] =\frac{1}{\pi} \int dq_0^{'}
exp(-\frac{q_0^{'2}} {M^2}) Im\Pi_+ (q_0^{'}). \label{BO}
\end{equation}
\end{subequations}
The r.h.s. of Eqs.~(\ref{BE}) and ~(\ref{BO}) are expressed in terms
of the observed spectral function. The absorptive part of the
projected correlation function is parameterized as
\begin{eqnarray}
&&Im\Pi_+ (q,p) = \bar u (q)i\gamma_5 u(q')\pi
\lambda^2g(q_0,\textbf{p}^2)[\frac{\delta
(q_0-m_N)}{q_0-E_k-\omega _p }\nonumber \\&&+\frac{\delta
(q_0-E_k-\omega _p)}{q_0-m_N}]+[\theta (q_0-s_\eta )+\theta
(-q_0-s_\eta)]\, Im\Pi_+^{OPE}(q,p),
\end{eqnarray}
where $s_\eta$  is the effective continuum threshold of $\eta N$
or $\eta \bar N$ channel and $m_N$  is the mass of the proton  and
$\lambda$ is the coupling of the proton current with one-proton state:
\begin{equation}
<0|J_N (0)|q>= \lambda u(q).
\end{equation}
We use the following parameterization of the vacuum-to-eta matrix
elements of the light-cone operators arising in our
calculation~\cite{Kim00,Doi04,Chen06,Kim08} $(q=u,d)$:
\begin{subequations}
\begin{equation}
<0|\bar q(0)i\gamma_5 q(x)|\eta(p)>= \frac{f_\eta \mu_\eta}{ \sqrt
6}[1-\frac{i}{2} p\cdot x-\frac{1}{6} (p\cdot x)^2],
\end{equation}
\begin{equation}
<0|\bar q(0) \gamma_\mu \gamma_5 q(x)|\eta(p)>=\frac{f_\eta}{
\sqrt 6}[ip_\mu+\frac{1}{2} (p\cdot x) p_\mu-\frac{i}{18} \delta
^2 p\cdot x x_\mu+\frac{5i}{36} \delta ^2 x^2 p_\mu + \frac{5}{72}
\delta ^2 p\cdot x x^2 p_\mu- \frac{ 1}{36} \delta ^2 (p\cdot x)^2
x_\mu],
\end{equation}
\begin{eqnarray}
<0|\bar q(0) \gamma_5 \sigma^{\mu\nu} q(x)|\eta(p)>=\frac{i}{6
\sqrt 6} (p^\mu x^\nu-p^\nu x^\mu ) f_\eta \mu_\eta
(1-\rho_+^\eta)(1-\frac{i}{2} p\cdot x),
\end{eqnarray}
\begin{eqnarray}
&&<0|g_s G_{\mu\nu}^n (\frac{1}{2} x) q_a (x) \bar q_b
(0)|\eta(p)>=\frac{if_{3\eta}}{16 \sqrt 6 }t_{ab}^n \gamma_5
p^{\lambda} (\sigma_{\lambda \mu} p_\nu-\sigma_{\lambda \nu}
p_\mu)\nonumber\\&&+\frac{if_\eta \delta ^2}{252 \sqrt 6} t_{ab}^n
[\frac{5}{8} \gamma_5 (\gamma_\mu p_\nu- \gamma_\nu p_\mu)px
+\gamma_5 \hat p (p_\mu x_\nu- p_\nu x_\mu)] \frac{f_\eta \delta
^2}{16 \sqrt 6} t_{ab}^n {\epsilon_{\mu\nu}}^{\alpha \beta
}[i\gamma_\alpha  p_\beta (\frac{1}{ 3}-\frac{i}{6} px) ].
\end{eqnarray}
\end{subequations}
The sign of  $\delta ^2$ is as per suggestion of Doi \emph{et
al}.~\cite{Doi04}. To leading order in SU(3)-flavor symmetry breaking, we
use the following Gell-Mann-Okubo-like relations for the
parameters of octet pseudo-scalar mesons $\pi$, $K$ and
$\eta$~\cite{Chen06,Kim08}:
\begin{subequations}
\begin{equation}
      3f_\eta+f_\pi =4f_K,
\end{equation}
\begin{equation}
3f_\eta \mu_\eta+f_\pi  \mu_\pi =4f_K \mu_K,
\end{equation}
\begin{equation}
3f_\eta \mu_\eta (1-\rho_+^\eta)+f_\pi  \mu_\pi  (1-\rho_-^\pi
)=4f_K  \mu_K (1-\rho_+^K ).
\end{equation}
\end{subequations}
The constants for pions and kaons are taken from Ref.~\cite{Ball}.
As a result, we get the following values for the constants for
$\eta$-meson:
\begin{eqnarray}
&&f_\eta=0.1695 GeV,\;\;\;\;\;\;\;\; \mu_\eta=1.6380 GeV,
\;\;\;\;\;\;\;\;f_{3\eta}=0.0045 GeV^2,
\;\;\;\;\;\;\;\;\rho_+^\eta=0.1028.
\end{eqnarray}
The decay constant f$\eta$ obtained is close to the value obtained
in a more elaborate scheme of two couplings and two mixing
angles~\cite{Singh}. The constant $\mu_\eta$ is also close to the
value directly obtained for $\eta$ as  $\mu_\eta = 3
m_\eta^2/(2\bar m+4m_s)$~\cite{Kim08}. The expression for the
correlation function obtained through operator product expansion
is: %%
\begin{eqnarray}
&&\Pi(q,p)=i\gamma_5 \hat p [\frac{f_\eta}{3 \sqrt 6 \pi ^2}
(q^2-\frac{1}{2} \delta ^2) ln(-q^2) +\frac{2}{9} \frac{f_\eta
\mu_\eta}{ \sqrt 6} <\bar q q> \frac{1}{q^2} -\frac{1}{36}
\frac{f_\eta \mu_\eta}{ \sqrt 6}<\bar q g_s \sigma\cdot Gq >
\frac{1}{q^4} \nonumber\\&&-<\frac{\alpha_s}{\pi} G^2> \frac{
f_\eta}{9 \sqrt 6} (\frac{1}{q^2} +\frac{2\delta ^2}{9q^4}) ]
-i\gamma_5 \frac{f_\eta \mu_\eta}{ \sqrt 6} [\frac{q^2}{4\pi ^2}
ln(-q^2) +\frac{1}{24q^2} <\frac{\alpha_s}{\pi}  G^2> ] -
i\gamma_5 \frac{3f_{3\eta}}{4 \sqrt 6 \pi ^2} p^2
ln(-q^2)\nonumber\\&&+\gamma_5 \sigma^{\mu \nu} q_\mu p_\nu
\frac{f_\eta \mu_\eta}{\sqrt 6}
 [\frac{1}{12\pi ^2}
ln(-q^2)-\frac{4}{3}\frac{<\bar q q>}{\mu_\eta} \frac{ 1}{q^2}
 -\frac{26}{27} \frac{<\bar q q>}{\mu_\eta}
\frac{\delta ^2}{q^4} +\frac{1}{2\mu_\eta} \frac{1}{q^4}<\bar q
g_s \sigma\cdot Gq>\nonumber\\&&+\frac{1}{216}<\frac{\alpha
_s}{\pi} G^2
>\frac{ 1}{q^4} ] +i\gamma_5 p\cdot q \hat q [-\frac{f_\eta}{3
\sqrt 6 \pi ^2} (ln (-q^2 )+\frac{7\delta ^2}{3q^2})+\frac{f_\eta
\mu_\eta}{ \sqrt 6}<\bar q q> \frac{ 4}{9} \frac{ 1}{q^4}
+\frac{f_\eta \mu_\eta}{ \sqrt 6}\frac{ 1}{9} <\bar q g_s
\sigma\cdot Gq> \frac{1}{q^6} \nonumber\\&&+\frac{8}{81}
\frac{f_\eta}{\sqrt 6} <\frac{\alpha _s}{\pi} G^2> \frac{\delta
^2}{q^6} -\frac{1}{18} \frac{f_\eta}{\sqrt 6}<\frac{\alpha_s}{\pi}
G^2>\frac{ 1}{q^4} ]+i\gamma_5 \frac{f_\eta}{ \sqrt 6} \rho_+^\eta
[\hat p \{-\frac{2\mu_\eta}{9} <\bar q q> \frac{ 1}{q^2}
+\frac{\mu_\eta}{36}<\bar q g_s \sigma \cdot Gq> \frac{1}{q^4}
\}\nonumber\\&&+ \hat q \{\frac{\mu_\eta}{18} <\bar q g_s \sigma
\cdot Gq> \frac{p^2}{q^6} +\frac{2\mu_\eta}{9} <\bar q q>\frac{
p^2}{q^4} \} ]+i\gamma_5 \frac{f_\eta}{\sqrt 6} p^2 \hat q
[-\frac{1}{6\pi ^2} ln(-q^2)+\frac{7\delta ^2}{18\pi^2 q^2}
-\frac{2\mu_\eta}{9}<\bar q q>   \frac{1}{q^4}
\nonumber\\&&+\frac{1}{36 q^4} <\frac{\alpha _s}{\pi}
G^2>-\frac{\mu_\eta}{18} <\bar q g_s \sigma\cdot Gq> \frac{1}{q^6}
+\frac{4}{81} <\frac{\alpha _s}{\pi} G^2> \frac{\delta ^2}{q^6}
]+i\gamma_5 \frac{f_\eta}{\sqrt 6} p\cdot q\nonumber\\&&[\hat p
\{\frac{1}{6\pi ^2} ln(-q^2)+ \frac{5\delta ^2}{9\pi ^2 q^2}
-\frac{\mu_\eta}{9} <\bar q g_s \sigma\cdot Gq> \frac{1}{q^6}
-\frac{1}{12} <\frac{\alpha _s}{\pi} G^2>(\frac{1}{q^4} -\frac{32
\delta ^2}{27q^6} )\}+ \hat q \frac{\mu_\eta}{9} <\bar q g_s
\sigma\cdot Gq> \frac{1}{q^6} ]\nonumber\\&&+i\gamma_5
\frac{f_\eta \mu_\eta}{ \sqrt 6} [\frac{p\cdot q}{4\pi ^2} ln(-q^2
)- \frac{p^2}{12\pi ^2} ln(-q^2)+\frac{p^2-3p\cdot q}{72q^4}
<\frac{\alpha _s}{\pi} G^2> ]+\gamma_5 \sigma^{\mu\nu} q_\mu p_\nu
\frac{f_\eta}{\sqrt 6} \nonumber\\&&\times[\rho_+^\eta
\{-\frac{\mu_\eta}{12\pi ^2} ln(-q^2)-\frac{\mu_\eta}{216}
<\frac{\alpha _s}{\pi}  G^2> \frac{1}{q^4} \}-\frac{20}{27}
\frac{\delta ^2}{q^4}   <\bar q q>+p\cdot
q\{-\frac{\mu_\eta}{12\pi ^2} \frac{1}{q^2} -\frac{4}{3} <\bar q
q> \frac{1}{q^4} -\frac{40}{27} <\bar q q> \frac{\delta ^2}{q^6}
\nonumber\\&&+<\bar q g_s \sigma\cdot Gq> \frac{1}{q^6}
+\frac{\mu_\eta}{108} <\frac{\alpha _s}{\pi} G^2> \frac{1}{q^6}
+\frac{4}{9} <\bar q q> \frac{\delta ^2}{q^6} \} ]. \label{ope}
\end{eqnarray}
In above expression, $q$  in $<\bar q q>$ and  $<\bar q g_s
\sigma\cdot Gq>$ stands for $u$ or $d$. The above expression has
some differences from the corresponding expressions obtained
previously by other authors~\cite{Kim00,Doi00}. We have given
complete expression for the mixed condensate term, and sign of
$\delta^2$, as pointed out earlier, is consistent with
Ref.~\cite{Doi04}. From Eq.~(\ref{ope}), we obtain  $\Pi_+^E
(q_0^2 )$ and $\Pi_+^O (q_0^2)$, as given in Eqs.~(\ref{PI}),
~(\ref{PE}) and ~(\ref{PO}). Using energy-momentum conservation,
it can be shown that $(k=q-p)$
\begin{equation}
\vec p ^2=-m_\eta^2+(m_\eta^4)/(4m_N^2),
\;\;\;\;\;\;\;\;\omega_p=(m_\eta^2)/(2m_N ),\;\;\;\;\;\;\;\; E_k
\cong m_N-\frac{m_\eta^2}{2m_N}.
\end{equation}
Upon Borel transform, $\Pi_+^E$  gives the following sum rule:
\begin{eqnarray}
&&2m_N^2 \lambda ^2 e^{- \frac{m_N^2}{M^2}} (g_{\eta
NN}-lM^2)=\nonumber\\&& -\frac{f_\eta \mu_\eta}{\sqrt 6}
[\frac{1}{\pi^2} M^6 E_1
(\frac{s_\eta}{M^2})(\frac{E_k+m_N-\omega_p}{3\mu_\eta}-\frac{1}{4})+
\frac{M^4}{\pi ^2} E_0 (\frac{s_\eta}{M^2})\{-\frac{5\delta ^2}{2}
\frac{(E_k+m_N-\omega _p)}{3\mu_\eta}-\frac{3}{4}\frac{
f_{3\eta}}{f_\eta \mu_\eta} m_\eta^2
\}\nonumber\\&&+(E_k+m_N-\omega _p)(\frac{2}{9} <\bar q
q>-\frac{1}{9} \frac{1}{\mu_\eta} <\frac{\alpha _s}{\pi}  G^2>)
M^2- \frac{1}{24} <\frac{\alpha _s}{\pi} G^2> M^2+(E_k+m_N-\omega
_p)(\frac{1}{36} <\bar q g_s \sigma\cdot Gq>\nonumber\\&&-
\frac{2}{81}\frac{\delta ^2}{\mu_\eta} <\frac{\alpha _s}{\pi}
G^2>) ]+\frac{f_\eta \mu_\eta}{ \sqrt 6} [\omega _p m_N^2
\{-\frac{M^4}{3\mu_\eta \pi ^2} E_0
(\frac{s_\eta}{M^2})+\frac{7\delta ^2 M^2}{9\pi ^2 \mu_\eta}
-\frac{4}{9} <\bar q q>+\frac{1}{18} \frac{ 1}{\mu_\eta}
<\frac{\alpha _s}{\pi} G^2>\nonumber\\&&+\frac{1}{18M^2} <\bar q
g_s \sigma\cdot Gq>- \frac{4}{81M^2} \frac{ \delta ^2}{\mu_\eta}
<\frac{\alpha _s}{\pi} G^2> \}+(E_k+m_N-\omega _p) \rho_+^\eta
\{\frac{2}{9} <\bar q q> M^2+  \frac{1}{36} <\bar q g_s
\sigma\cdot Gq> \}\nonumber\\&&+m_\eta^2 \{\frac{M^4}{12\pi ^2}
E_0 (\frac{s_\eta}{M^2})+ \frac{1}{72} <\frac{\alpha _s}{\pi} G^2>
\}+ \omega_p (E_k+m_N)\{-\frac{4}{3} <\bar q q> \frac{1}{\mu_\eta}
M^2+\frac{28}{27} \frac{\delta ^2}{\mu_\eta}  <\bar q
q>-\frac{1}{\mu_\eta} <\bar q g_s \sigma\cdot Gq>\nonumber\\&&-
\frac{1}{108} <\frac{\alpha _s}{\pi}  G^2> \} ]. \label{PESR}
\end{eqnarray}

On the l.h.s. $l$ is a constant independent of $M^2$. On the r.h.s.,
$E_0(x) = 1-e^{-x}$ and $E_1(x) = 1-(1+x)e^{-x}$ are used to model
contributions of excited states. The expression in the second
square bracket on the r.h.s. in Eq.~(\ref{PESR}) gives purely
SU(3) symmetry breaking contribution. After Borel transform
$\Pi_+^O$ gives the following sum rule
\begin{eqnarray}
&&2m_N \lambda ^2 e^{- \frac{m_N^2}{M^2}} (g_{\eta
NN}-hM^2)\nonumber\\&&=\frac{f_\eta \mu_\eta}{ \sqrt 6}
[\frac{M^4}{12\pi ^2} (E_k+m_N) E_0
(\frac{s_\eta}{M^2})-\frac{4}{3} \frac{<\bar q q>}{\mu_\eta}
(E_k+m_N) M^2+\{\frac{26}{27}\frac{<\bar q q>}{\mu_\eta} \delta
^2-\frac{1}{2} \frac{m_0^2 <\bar q q>}{\mu_\eta}
\nonumber\\&&-\frac{1}{216} <\frac{\alpha _s}{\pi} G^2> \}(E_k+m_N
) ]+\frac{f_\eta \mu_\eta}{ \sqrt 6}
[-\{\frac{2m_\eta^2}{\mu_\eta} +\frac{2\omega _p}{\mu_\eta}
(E_k+m_N- \omega _p)+3\omega _p+\rho_+^\eta (E_k+m_N) \}
\frac{M^4}{12\pi ^2} E_0
(\frac{s_\eta}{M^2})\nonumber\\&&+\{\frac{7}{18\pi ^2} \delta ^2
\frac{m_\eta^2}{\mu_\eta} -\frac{5}{9\pi ^2} (E_k+ m_N-\omega
_p)\delta ^2 \frac{\omega _p}{\mu_\eta}\} M^2-\frac{1}{36}
\frac{m_\eta^2}{\mu_\eta}<\frac{\alpha _s}{\pi} G^2>-\frac{2}{9}
m_\eta^2 <\bar q q>(\rho_+^\eta-1)\nonumber\\&&-\{\frac{1}{12}
\frac{\omega _p}{\mu_\eta} (E_k+m_N-\omega _p)+\frac{\omega
_p}{24}-\frac{\rho_+^\eta (E_k+m_N)}{216}\} <\frac{\alpha _s}{\pi}
G^2>+\rho_+^\eta (E_k+ m_N) \frac{ 20}{27} \frac{\delta
^2}{\mu_\eta} <\bar q q>+\{(\rho_+^\eta-1)
\frac{m_\eta^2}{36}\nonumber\\&&+\frac{\omega _p}{18}
(E_k+m_N-\omega _p) \} \frac{m_0^2}{M^2} <\bar q q>+ \frac{2}{81}
\frac{\delta ^2}{\mu_\eta} <\frac{\alpha _s}{\pi}
G^2>\{m_\eta^2-2(E_k+m_N-\omega _p) \omega _p \} \frac{1}{M^2} ].
\label{POSR}
\end{eqnarray}
In Eq.~(\ref{POSR}) also, the expression in the second square
bracket on the r.h.s. gives purely SU(3) symmetry breaking
contribution. On the l.h.s., $h$ is a constant independent of
$M^2$. We also write the nucleon mass sum rules~\cite{Singh,Ioffe}
for our use. The chiral-even mass sum rule is
\begin{equation}
m_N \lambda ^2 e^{- \frac{m_N^2}{M^2}}= \frac{1}{4\pi ^2}  [-M^4
E_1 (\frac{s_0}{M^2})<\bar q q>+\frac{5\pi ^2}{18} <\bar q
q><\frac{\alpha _s}{\pi} G^2> ].
\end{equation}
In above equation, $s_0$ is the continuum threshold for the mass
sum rule. The coefficient of dimension 7 operator has been  taken
from Ref.~\cite{Ioffe}. The chiral-odd mass sum rule is
\begin{equation}
\lambda ^2 e^{- \frac{m_N^2}{M^2}}=\frac{1}{4\pi ^2}
[\frac{M^6}{8} E_2 (\frac{s_0}{M^2})+\frac{M^2}{8} E_0
(\frac{s_0}{M^2}) \pi ^2 <\frac{\alpha _s}{\pi} G^2>+\frac{8\pi
^4}{3} <\bar q q>^2-\frac{2\pi^4}{3}  <\bar q g_s \sigma\cdot
Gq><\bar q q> \frac{1}{M^2} ]. \label{ioffe}
\end{equation}

In Eq.~(\ref{ioffe}), $E_2(x)=1-(1+x+x^2/2)e^{-x}$. By dividing the
two coupling sum rules with the two mass sum rules we get a total
of four equations, each of which can be used independently for
the determination of  the coupling $g_{\eta NN}$. We will study full
sum rules as well as SU(3) symmetric part of sum rules in the next
section.

\section{ANALYSIS OF  SUM RULES AND RESULTS}
\label{ana}
\begin{figure}[hb]
\centerline{\psfig{file=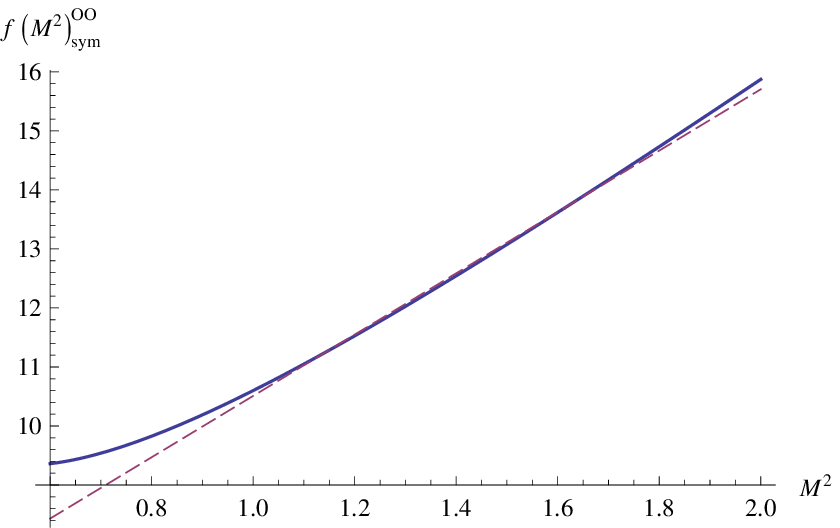,width=4.2in}} \vspace*{-0cm}
\caption{\small{Plot of leading order  terms in the odd-odd sum
rule as a function of $M^2$ (solid curve). A straight line fit of
the form  $5.20 M^2 + 5.31$ (broken) over the range 1.0 GeV$^2$
$\le M^2 \le 1.8$ GeV$^2$   gives  $g_{\eta NN}=5.31$.}}
\label{fig1}
\end{figure}

\begin{figure}[hb]
\centerline{\psfig{file=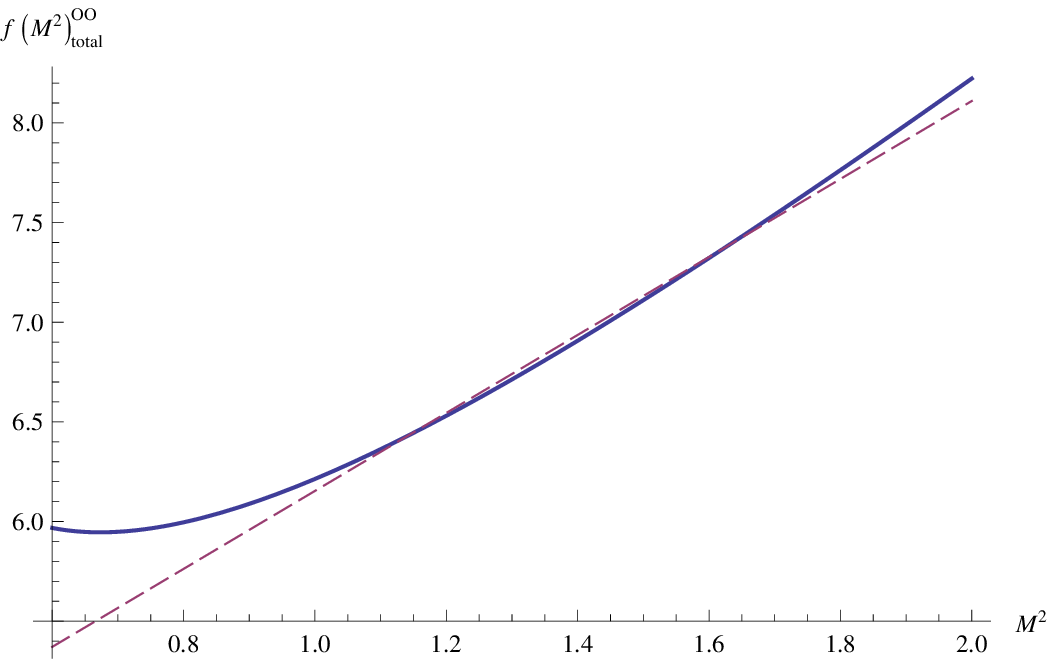,width=4.2in}} \vspace*{-0cm}
\caption{\small{Plot of full expression obtained in the odd-odd
sum rule as a function of $M^2$ (solid curve). A straight line fit
of the form  $1.96 M^2  + 4.20$ (broken) over the range 1.0
GeV$^2$ $\le M^2 \le 1.8 $ GeV$^2$   gives  $g_{\eta NN}=4.20$.
$\chi^2$ (defined in the text) for this fit is $2.0 \times
10^{-3}$.}} \label{fig2}
\end{figure}

\begin{figure}
\centerline{\psfig{file=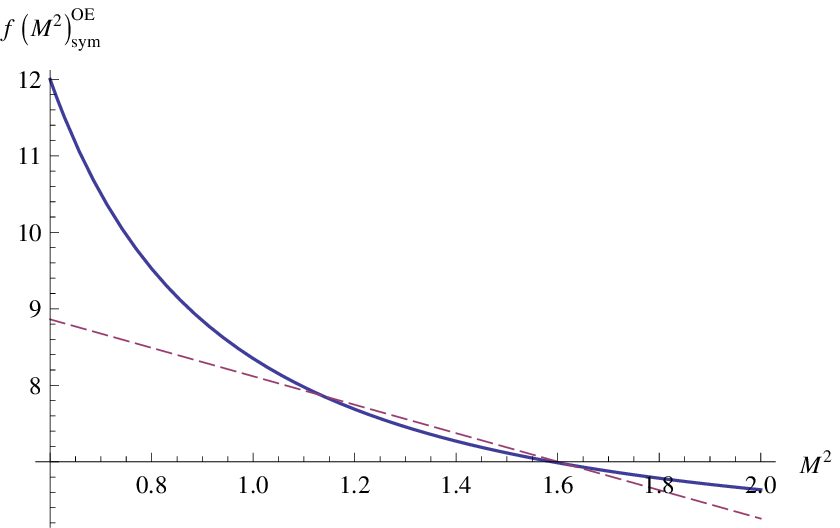,width=4.2in}} \vspace*{-0cm}
\caption{\small{Plot of leading order  terms in the odd-even sum
rule as a function of $M^2$ (solid curve). A straight line fit of
the form $-1.86 M^2 + 9.98$ (broken) over the range 1.0 GeV$^2$
$<M^2 <1.8$ GeV$^2$  gives  $g_{\eta N N}$=9.98.}} \label{fig3}
\end{figure}

\begin{figure}
\centerline{\psfig{file=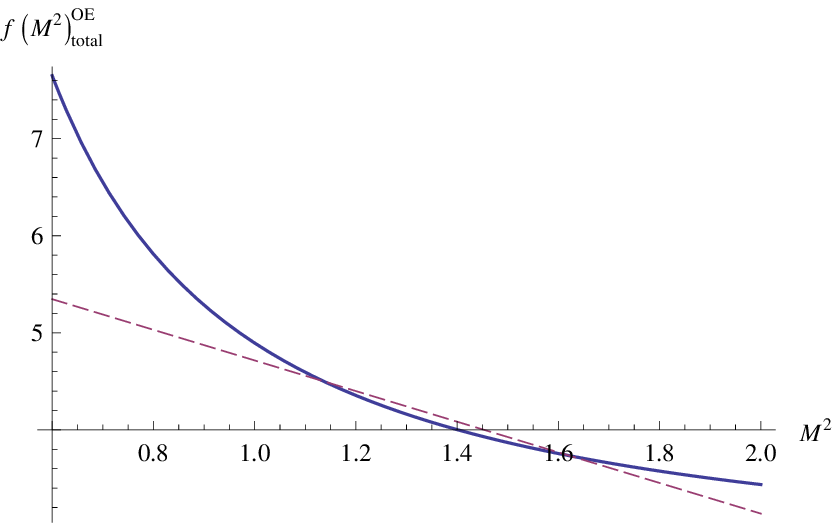,width=4.2in}} \vspace*{-0cm}
\caption{\small{Plot of full  expression obtained in the odd-even
sum rule as a function of $M^2$ (solid curve). A straight line fit
of the form $-1.58 M^2 +6.29$ (broken curve) over the range 1.0
GeV$^2$ $\le M^2\le 1.8$ GeV$^2$ gives $g_{\eta N N}=6.29$.
$\chi^2$ (defined in the text) for this fit is $9.2\times 10^{-3}$.
A better fit is shown in Fig.~\ref{fig5}.}} \label{fig4}
\end{figure}

\begin{figure}
\centerline{\psfig{file=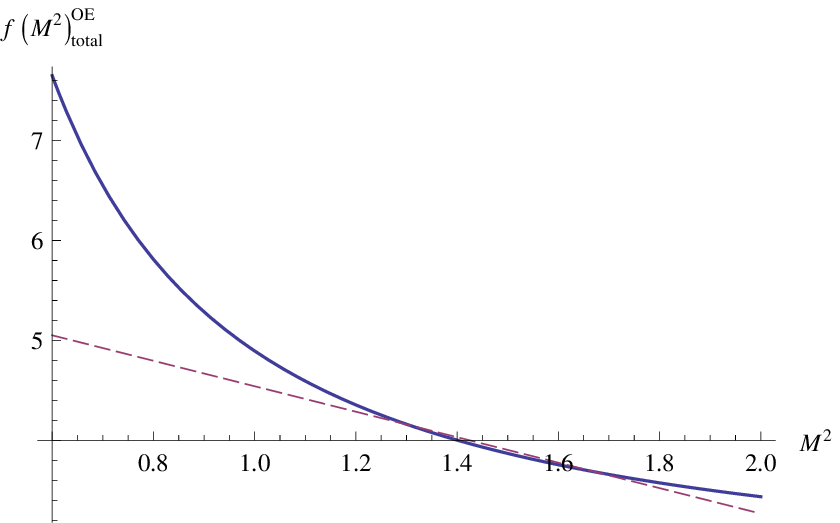,width=4.2in}} \vspace*{-0cm}
\caption{\small{Plot of full  expression obtained in the odd-even
sum rule as a function of $M^2$ (solid curve). A straight line fit
of the form $-1.27 M^2 +5.82$ (broken curve) over the range 1.2
GeV$^2$ $\le M^2\le 1.8$ GeV$^2$ gives $g_{\eta N N}$=5.82.
$\chi^2$ for this fit is $4.0\times 10^{-3}$.}} \label{fig5}
\end{figure}

\begin{figure}
\centerline{\psfig{file=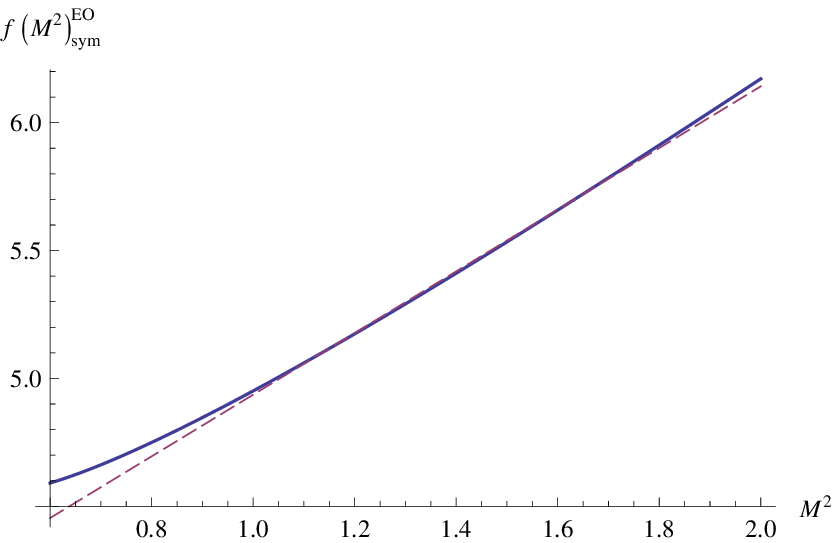,width=4.2in}} \vspace*{-0cm}
\caption{\small{Plot of leading order  terms in the even-odd sum
rule as a function of $M^2$ (solid curve). A straight line fit of
the form $1.21M^2 + 3.73$ (broken curve) over the range 1.0
GeV$^2$  $\le M^2 \le 1.8$ GeV$^2$ gives $g_{\eta N N}=3.73$.}}
\label{fig6}
\end{figure}

\begin{figure}
\centerline{\psfig{file=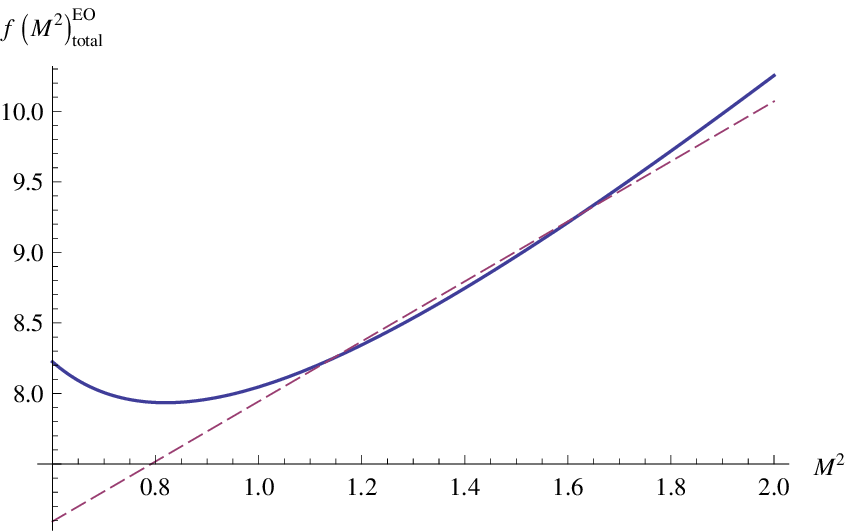,width=4.2in}} \vspace*{-0cm}
\caption{\small{Plot of full  expression obtained in the even-odd
sum rule as a function of $M^2$ (solid curve). A straight line fit
of the form $2.13 M^2 + 5.82$ (broken curve) over the range 1.0
GeV$^2$  $\le M^2 \le 1.8$ GeV$^2$ gives $g_{\eta N N}=5.82$.
$\chi^2$ for this fit is $2.61\times 10^{-3}$.}} \label{fig7}
\end{figure}

\begin{figure}
\centerline{\psfig{file=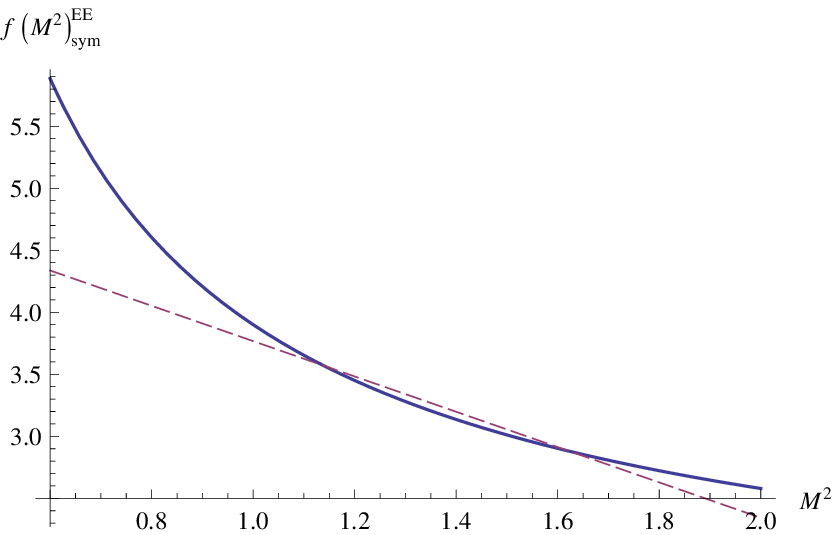,width=4.2in}} \vspace*{-0cm}
\caption{\small{Plot of leading order  terms in the even-evn sum
rule as a function of $M^2$ (solid curve). A straight line fit of
the form $-1.42 M^2 +5.19$ (broken curve) over the range 1.0
GeV$^2$ $\le M^2 \le 1.8$ GeV$^2$ gives $g_{\eta N N}=5.19$.}}
\label{fig8}
\end{figure}

\begin{figure}
\centerline{\psfig{file=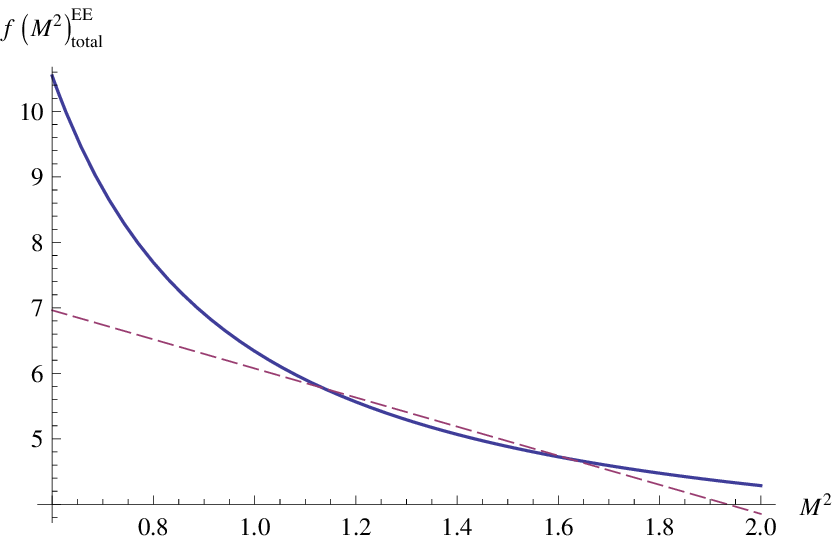,width=4.2in}} \vspace*{-0cm}
\caption{\small{Plot of full  expression obtained in the even-even
sum rule as a function of $M^2$ (solid curve). A straight line fit
of the form  $-2.22 M^2 +8.30$ (broken curve) over the range 1.
0 GeV$^2$ $\le M^2\le 1.8$ GeV$^2$ gives $g_{\eta N N}=8.30$.
$\chi^2$ for this fit is $10.6\times 10^{-3}$. A better fit is
shown in Fig.~\ref{fig10}.}} \label{fig9}
\end{figure}

\begin{figure}
\centerline{\psfig{file=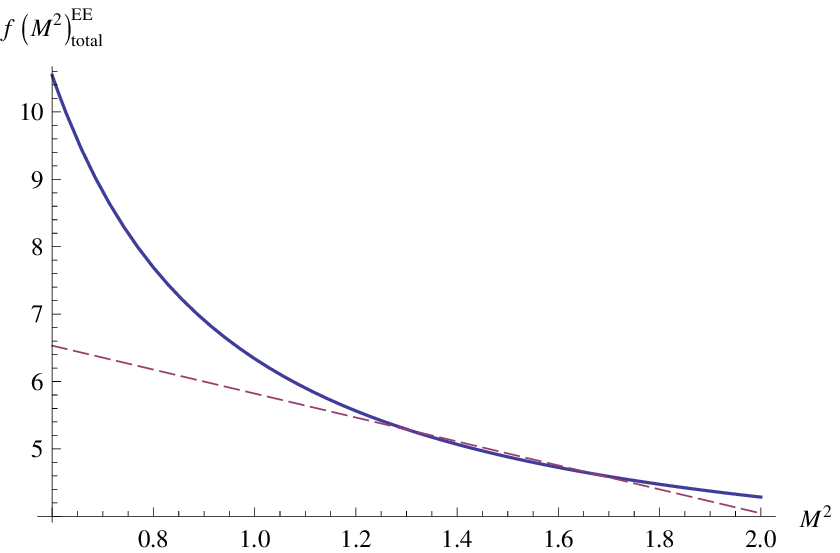,width=4.2in}} \vspace*{-0cm}
\caption{\small{Plot of full  expression obtained in the even-even
sum rule as a function of $M^2$ (solid curve). A straight line fit
of the form  $-1.78 M^2 +7.60$ (broken curve) over the range 1.
2 GeV$^2$ $\le M^2\le 1.8$ GeV$^2$ gives $g_{\eta N N}=7.60$.
$\chi^2$ for this fit is $4.5\times 10^{-3}$.}} \label{fig10}
\end{figure}

\begin{figure}
\centerline{\psfig{file=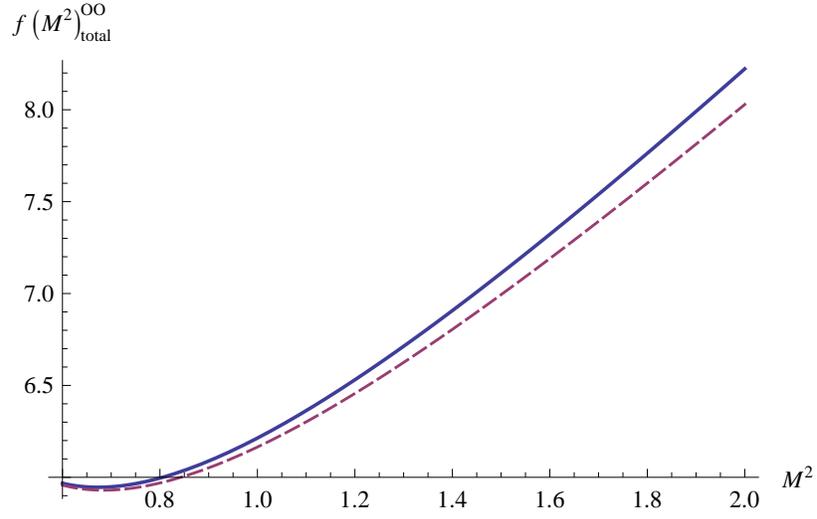,width=4.2in}} \vspace*{-0cm}
\caption{\small{Plot of full expression for odd-odd sum rule for
$s_\eta=2.57$ GeV$^2$ (solid curve) and $s_\eta=2.07$ GeV$^2$
(broken curve) as a function of $M^2$. This changes $g_{\eta N N}$
from 4.20 to 4.29.}} \label{fig11}
\end{figure}

\begin{figure}
\centerline{\psfig{file=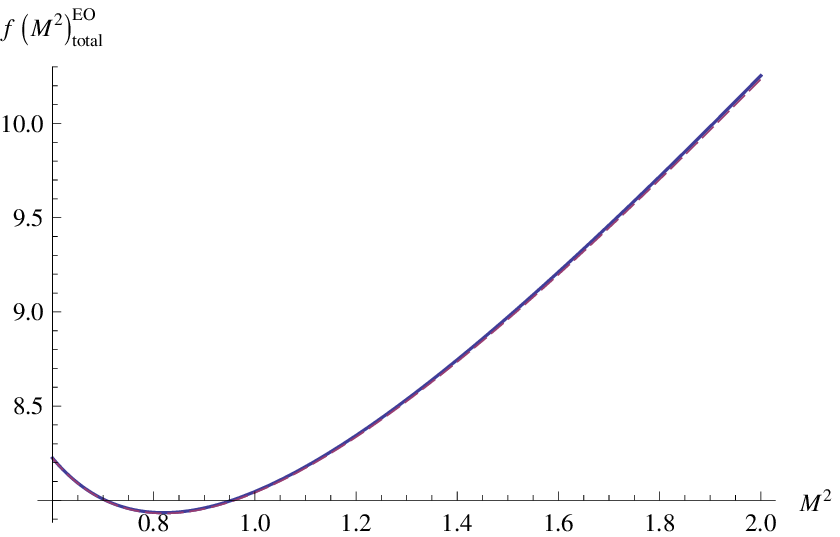,width=4.2in}} \vspace*{-0cm}
\caption{\small{Plot of full expression for even-odd sum rule for
$s_\eta=2.57$ GeV$^2$ (solid curve) and $s_\eta=2.07$ GeV$^2$
(broken curve) as a function of $M^2$. This does not change
$g_{\eta N N}$ to the accuracy we are working.}} \label{fig12}
\end{figure}

We have used the following numerical constants in our analysis
\begin{eqnarray}
&&<\bar q q> = - (0.23\;GeV)^3,\;\;\;\;\;\;\;\;  <\frac{\alpha
_s}{\pi} G^2>=(0.33 \;GeV)^4, \;\;\;\;\;\;\;\;  <\bar q g_s
\sigma\cdot Gq> =m_0^2 <\bar q q> \;\;\;\; \nonumber \\&& with\;\;
m_0^2=0.8 GeV^2,\;\;\;\;\;\;\;\; f_{3\eta}=0.0045, m_N=0.939\;
GeV, \;\;\;\;\;\;\;\; m_\eta=0.547 \;GeV, \;\;\;\;\;\;\;\; \delta
^2=0.2 \;GeV^2.
\end{eqnarray}
For continuum threshold  $s_\eta$ in the coupling sum rules we
have uniformly used $s_\eta = 2.57$ GeV$^2$ in both coupling constant sum rules
while $s_0= 2.00$ GeV$^2$ in both mass sum rules. Later on we
shall study the effect of changes in the continuum thresholds on
results obtained for $g_{\eta N N}$. We use a notation in which
$f_{total}^{EO}$ is the function obtained by taking the ratio of
full OPE side of even sum rule in coupling to odd sum rule in mass
while $f_{sym}^{EO}$ has a similar meaning except that in coupling
sum rule only SU(3) symmetric terms are taken, and similarly for
others. We have shown the plots of all the functions obtained by
taking the ratio of a coupling sum rule to a mass sum rule.
Furthermore, for coupling sum rules both the full expressions as
given in Eqs.~(\ref{PESR}) and ~(\ref{POSR}) as well as SU(3)
symmetric parts have been considered. We have found straight line
fits for all the curves over a range of Borel mass $1.0$ GeV$^2$
$\le M^2 \le 1.8$ GeV$^2$. To get a quantitative idea of the
goodness of fit, we define a $\chi^2$ by
\begin{equation}
\chi^2=[\sum_{i=0}^n \{f(x_i)-f_{fit} (x_i) \}^2/ \{f(x_i)+f_{fit}
(x_i ) \}^2 ]/(1+n).  \label{chi}
\end{equation}
All the curves obtained from ratios and their straight line fits
are shown in Figs.~\ref{fig1} to~\ref{fig12}. The Borel windows
for both coupling sum rules are common, since, in principle,
couplings for all the Lorentz structures are related under SU(3)
rotations~\cite{Doi00}. We have listed the results of fits along
with $\chi^2$'s in Table~\ref{genn}. It is observed that ratios
containing even mass sum rule, namely, $f_{total}^{OE}$ and
$f_{total}^{EE}$, exhibit poor fits. Hence, we have also shown
their fits over a shorter Borel window ( 1.2 GeV$^2 \le M^2 \le
1.8 $ GeV$^2$) in Figs.~\ref{fig5} and~\ref{fig10}. This gives
better $\chi^2$, but still not so good as that for $f_{total}^{OO}$
and $f_{total}^{EO}$. Hence, hereafter we analyze these two latter
sum rules.

%
%%%%%%%%%%%%%%%%%%%%%%%%%%%%%%%%%%%%%%%%%%%%%%%%%%%%%%%
\begin{table*}[thb] % * for wide table
\caption{Numerical values obtained for  $g_{\eta N N}$ from
various sum rules and the $\chi^2$ (defined in Eq.~(\ref{chi})) for
straight line fit over the Borel window given in second row. For
all the cases $n=20$ has been used to get $\chi^2$.} \label{genn}
 \begin{tabular*}{0.92\textwidth}{@{\extracolsep{\fill}}lcccccc}
\hline\hline
   Sum rule &   Even-odd& Odd-odd &Even-even  &  Odd-even& Even-even &Odd-even
\\\hline Range of $M^2$ (GeV$^2$) &1.0 - 1.8  &  1.0 - 1.8 &1.
0 - 1.8 &1.0 - 1.8  & 1.2- 1.8 &1.2 - 1.8
 \\  $g_{\eta N N}$ in SU(3) sym. limit  &  3.73  &  5.31 &5.19  &  9.98  &  4.83 &9.36
 \\$g_{\eta N N}$for full expression &5.82   & 4.20 &   8.30 &6.29& 7.60 &5.82
 \\$\chi^2$  for st. line fit of  $f_{total}$  &2.61$\times$ 10$^{-3}$ &1.98$\times$ 10$^{-3}$ &10.6$\times$ 10$^{-3}$ &9.23$\times$ 10$^{-3}$ &  4.55$\times$ 10$^{-3}$  & 4.00$\times$ 10$^{-3}$
\\
\hline\hline
\end{tabular*}
\end{table*}

%%%%%%%%%%%%%%%%%%%%%%%%%%%%%%%%%%%%%%%%%%%%%%%%%%%%%%%
%
It is to be noted that $\Pi_+^E$  contains combination of SU(3)
symmetric axial vector and pseudo-scalar terms whereas  $\Pi_+^O$
contains tensor terms in SU(3) symmetry limit. The use of tensor
structure has been advocated on account of its nice features. It
has been concluded that for the later the physical parameter, in
SU(3) symmetry limit, is independent  of the form of the nucleon
current for currents with no derivative~\cite{Doi00}, the sum rule
does not depend on the form of the coupling in the effective
Lagrangian~\cite{Kim99}, Borel curves are rather insensitive to
the continuum threshold~\cite{Doi00}, etc. In the present case, we
see from Table~\ref{genn} that the odd-odd sum rule has the lowest
$\chi^2$. We have found that the contribution of excited states to
$\Pi_+^O$ is $6.5\%$ at $M^2=1.8\%$ GeV$^2$ while the operators
of highest  dimension contribute $6.7\%$  at $M^2=1.0$ GeV$^2$.
 Hence, the Borel window is quite appropriate for this case. The change
 of continuum threshold $s_\eta$ from $2.57$  GeV$^2$ to $2.07$
 GeV$^2$ (see Fig.~\ref{fig11})
 increases the value of $g_{\eta N N}$ by $2.3\%$ while the change of
 the continuum threshold $s_0$ from $2.0$  GeV$^2$ to $2.3$  GeV$^2$
  increases it by $3.7\%$. Change of  $<\bar q q>$ in both the coupling
  and mass sum rules of  $f_{total}^{OO}$ as $-(0.23\pm 0.025$ GeV)$^3$
   changes  $g_{\eta N N}$  by $9.6\%$. Uncertainties of gluon condensate,
   $m_0^2$  and $\delta ^2$ may change the value of $g_{\eta N N}$ by up to $5\%$.
    The uncertainties in  $f_\eta$,  $\mu_\eta$, $f_{3\eta}$ and $\rho_+^\eta$
     (which are small);  higher powers of $m_\eta$  not considered and
      extrapolation of the coupling constant to the physical point  may
       result in another couple of percent of change in $g_{\eta N N}$.
       In all, we expect an uncertainty of up to $25\%$ in the value of
         $g_{\eta N N}$ obtained from analysis of $f_{total}^{OO}$ due to
          uncertainties in various constants used in the sum rule and
          extrapolation.

The even-odd sum rule has $\chi^2$  marginally higher than the odd-odd case.
 Also, the value of $g_{\eta N N}$  obtained from even-odd sum rule coincides
 with the one obtained from odd-even sum rule, albeit with a lower Borel window
  ($1.2 - 1.8$ GeV$^2$). Moreover, this sum rule has a minimal dependence on
  the continuum threshold   $s_\eta$, as is evident from Fig.~\ref{fig12}.
  However, contribution of excited states as per the model used here is $23\%$ at
  $M^2 =1.8$  GeV$^2$. Highest dimensional operators contribute $24\%$ to
  $\Pi_+^E$ in the entire range of Borel window. This means that the omitted
  higher order terms in OPE may contribute significantly to $\Pi_+^E$.
  Changing $<\bar q q>$ to $-(0.255$ GeV)$^3$ in both the coupling and the mass sum
  rule of  $f_{total}^{EO}$ changes  $g_{\eta N N}$  by $23\%$. 
    This means that the result obtained from even-odd sum rule has a larger uncertainty.
     Hence, we consider the result obtained from odd-odd sum rule as the most reliable
     and as our main result we have:
\begin{equation}
 g_{\eta N N} = 4.20 \pm 1.05
\end{equation}
A few comments on the size of SU(3)-breaking effects in the four sum rules. 
For even-odd and even-even sum rules, the coupling $g_{\eta N N}$ increases 
on introducing symmetry breaking effects and the increments (with respect to the SU(3) 
symmetric results) are almost uniform (by 56 to 60\%) in these cases. 
On the other hand, the coupling decreases for odd-odd and odd-even cases on introducing symmetry breaking effects. Furthermore, the change is minimum (about 21\%) for odd-odd sum rule 
and moderate (about 38\%) for odd-even case.

In Table~\ref{literature}, we compare our result with the values
of $g_{\eta N N}$ obtained by other authors using various methods
in the recent literature.
In Refs.~\cite{Feldmann,Shore,Nasrallah} Goldberger-Treiman relation has been used to calculate 
$g_{\eta N N}$ and $g_{\eta' N N}$ simultaneously. 
Authors of Refs. [8,9] use the measured values of singlet and octet axial charges 
of nucleon. 
Nasrallah~\cite{Nasrallah}, on the other hand, has used dispersion relation for the nucleon matrix 
element of the divergence of axial current saturating it with $\eta$ and $\eta'$ states 
and accounting the continuum contribution through a model. 
Feldmann~\cite{Feldmann} has neglected the contribution of higher excited pseudo-scalar states. 
In the absence of knowledge of gluon topological susceptibility and gluonic coupling 
to nucleon ($g_{GNN}$), Shore [9] ends up getting a plot of  
$g_{\eta N N}$ and $g_{G N N}$ as a function of $g_{\eta' N N}$.
Kim {\it et. al.}~\cite{Kim00} have performed calculation of $g_{\eta N N}$  to leading order 
in $p_\mu$ by considering the two separate Dirac structures $i\gamma_5\gamma_\mu p^\mu$ 
and and $\gamma_5\sigma_{\mu\nu} q^\mu p^\nu$ separately. 
They have introduced SU(3)-breaking only through the decay constant $f_\eta$. 
In Ref.~\cite{Stoks}, while fitting the data in a potential model that accounts for various meson 
exchanges, the coupling $g_{\eta N N}$ was not searched independently; 
rather it was constrained via SU(3) symmetry.
In Ref.~\cite{Benmerrouche} the authors have worked with effective Lagrangian approach to calculate  
$\eta$ photoproduction in the N*(1535) region. 
They find that the observed differential cross section is not very sensitive 
to either the nature of the $\eta$-nucleon coupling or to the precise value of the 
coupling constant. 
They extract a broad range of values for the $\eta NN$ pseudo-scalar coupling constant: 
$0.2 \le g_{\eta N N} \le 6.2$ from analysis of all available data.

%
%%%%%%%%%%%%%%%%%%%%%%%%%%%%%%%%%%%%%%%%%%%%%%%%%%%%%%%
\begin{table*}[thb] % * for wide table
\caption{Some of the results for eta-nucleon coupling obtained
in recent literature.} \label{literature}
 \begin{tabular*}{0.99\textwidth}{@{\extracolsep{\fill}}lcccccccc}
\hline\hline
 Ref. no. &~\cite{Nasrallah}    &  ~\cite{Feldmann}   &  ~\cite{Kim00} &  ~\cite{Stoks}  & ~\cite{Dumbrajs} &  ~\cite{Shore}    & ~\cite{Benmerrouche}  & This work
 \\\hline $g_{\eta N N}$  & 5.20$\pm$0.25   & 3.4$\pm$0.5    &   5.76&
   7.95  & 6.8     & 3.96 $\pm$ 0.16 & 3.2$\pm$3.0  &4.20$\pm$ 1.05 \\
&&&[SU(3) sym ]&&&($g_{\eta'NN}=2.0$)&&\\ &&&7.34&&& 3.59$\pm$0.15&&\\
&&&[Brok. SU(3)]&&& ($g_{\eta'NN}=1.0$)&&\\\hline
  Specification &   GT relation&  GT relation & Sum
rule&  Pot. model & Pot. model & GT relation & Expt. &Sum rule
\\
\hline\hline
\end{tabular*}
\end{table*}

\section{SUMMARY AND CONCLUSION}
\label{con}

Eta-nucleon coupling constant has a definite role in analyzing
experiments involving NN scattering in general and $\eta$
production in particular. However, due to the dominant role of
N*(1535) in $\eta$ production and lack of precise knowledge of
$g_{\eta N N}$, the role of $\eta$NN vertex is usually ignored in
analysis~\cite{Machleidt,Shyam}. Hence a credible determination of
$g_{\eta N N}$ is desirable. Separating the even and odd parts in
the projected correlation function, we obtained two independent
sum rules for the coupling $g_{\eta N N}$. These sum rules
incorporate the effect of SU(3)-flavor symmetry breaking in leading
order. The symmetry breaking parameters themselves come from
matrix elements of quark and gluon operators between vacuum and
one-eta state. Taking the ratios of the two coupling sum rules
with the two mass sum rules for the nucleon, we obtained a total of
four sum rules. We have analyzed the full expression for the four
cases as well as their SU(3) symmetric forms. The results obtained
from three of these full sum rules are quite close to each other
and this enhances the credibility of our result. The result from
odd-odd sum rule is quite robust and is the main result of this
work.

We have ignored the mixing of octet state with singlet and
glueball states and the consequent role of instantons in our
analysis. Nevertheless, we have quantified the SU(3) breaking
corrections to eta-nucleon coupling constant to leading order in
SU(3) symmetry breaking, and found the effects to be significant. 
We conclude that inclusion of SU(3)
symmetry breaking effects in a realistic way in this work is a
step forward in realizing a correct estimate of this poorly
determined quantity.

\begin{acknowledgments}
This work is supported in part by U.S. Department of Energy under grant
DE-FG02-95ER-40907.
J.P.S. thanks GWU for the hospitality where part of the
work was done, and DST, New Delhi for providing financial assistance.
\end{acknowledgments}

%%%%--------------------------------------------------------------------------------
%%%%--------------------------------------------------------------------------------

\begin{thebibliography}{00}
%
\bibitem{Machleidt} R. Machleidt, Phys. Rev. C \textbf{63}, 024001 (2001).

\bibitem{Ceci} S. Ceci, A. Svarc and B. Zauner, arXiv:0904.2430.

\bibitem{Stoks} V. G. J. Stoks and T. A. Rijken, Phys. Rev. C \textbf{59}, 3009 (1999);
 T.A. Rijken, Phys. Rev. C \textbf{73}, 044007 (2006).

\bibitem{Halperin} I. Halperin, Phys. Rev. D \textbf{50},4602 (1994).

\bibitem{Nolen} J. A. Nolen and J. P. Schiffer, Ann. Rev. Nucl. Sci. \textbf{19}, 471
(1969); G. A. Miller, B. M. K. Nefkens and I. Slaus, Phys. Rep.
\textbf{194}, 1 (1990).


\bibitem{Ellis} 
J. Ellis, hep-ph/0411369; J.R. Ellis, Nucl. Phys. \textbf{A684}, 53  (2001).

\bibitem{Sullivan}J. D. Sullivan, Phys. Rev. D \textbf{5}, 1732.

\bibitem{Feldmann} T. Feldmann, Int. J. Mod. Phys.  \textbf{A15}, 159 (2000).

\bibitem{Shore} G. M. Shore, Nucl. Phys. \textbf{B744}, 34 (2006).

\bibitem{Hanhart} C. Hanhart, arXiv: nucl-th/0511045 and references therein.

\bibitem{Reinders85} L. J. Reinders, H. R. Rubinstein and S. Yazaki, Phys. Rep.
\textbf{127}, 1 (1985).

\bibitem{Reinders84} L. J. Reinders, Acta Phys. Pol. \textbf{B15}, 329 (1984).

\bibitem{Birse} M. C. Birse and B. Krippa, Phys. Lett. \textbf{B373}, 9 (1996).

\bibitem{Kim00} H. Kim, T. Doi, M. Oka and S. H. Lee, Nucl. Phys. \textbf{A678}, 295
(2000).

\bibitem{Doi00} T. Doi, H. Kim and M. Oka, Phys. Rev. C \textbf{62}, 055202 (2000).

\bibitem{Kim99} H. Kim, S. H. Lee and M. Oka, Phys. Rev. D \textbf{60}, 034007 (1999).

\bibitem{Kondo02} Y. Kondo and O. Morimatsu, Phys. Rev. C \textbf{66}, 028201 (2002).

\bibitem{Kondo03} Y. Kondo and O. Morimatsu, Nucl. Phys. \textbf{A717}, 55 (2003).

\bibitem{Doi04} T. Doi, Y. Kondo and M. Oka, Phys. Rep. \textbf{398}, 253 (2004).

\bibitem{Aliev} T. M. Aliev \emph{et al}., Phys. Rev. D \textbf{74}, 116001 (2006).

\bibitem{Belyaev} V. M. Belyaev, V. M. Braun, A. Khodjamirian and R. Ruckel,
Phys. Rev. D \textbf{51}, 6177 (1995).

\bibitem{Ball99} P. Ball, JHEP \textbf{9901}, 010 (1999).

\bibitem{Ball} P. Ball, V. M. Braun and A. Lenz, arXiv: hep-ph/0603063 and
references therein.

\bibitem{Chen06} J-W Chen, H-M Tsai and K-C Weng, Phys. Rev. D \textbf{73}, 054010
(2006).

\bibitem{Kim08} C. Kim and A. K. Leibovich, Phys. Rev. D \textbf{78}, 054026 (2008).


\bibitem{Singh} Janardan P. Singh and J. Pasupathy, Phys. Rev. D \textbf{79}, 116005
(2009).

\bibitem{Ioffe} B. L. Ioffe, Acta Phys. Polon. \textbf{B16}, 543 (1985).

\bibitem{Leinweber} D. B. Leinweber, Ann. Phys. \textbf{254}, 328 (1997); ibid 198, 203
(1990).

\bibitem{Nasrallah} N. F. Nasrallah,, Phys. Lett. \textbf{B645}, 335 (2007).

\bibitem{Dumbrajs} O. Dumbrajs \emph{et al}., Nucl. Phys. \textbf{B216}, 277 (1983) and references therein.

\bibitem{Benmerrouche} M. Benmerrouche, N. C. Mukhopadhyay and J. F. Zhang, Phys.
Rev. D \textbf{51}, 3237 (1995).

\bibitem{Shyam} R. Shyam, Phys. Rev. C \textbf{75}, 055201 (2007).

\end{thebibliography}
\end{document}